\documentclass[prl,aps,prstab,fleqn,preprint,superscriptaddress,groupedaddress]{revtex4-1}

\usepackage{bm}
\usepackage{graphicx}
\usepackage{adjustbox}
\usepackage{amsmath}
\usepackage{MnSymbol}

\begin{document}
\title{Nanoscale magnetization and current imaging using scanning-probe magneto-thermal microscopy}

\author{Chi Zhang}
\author{Jason M. Bartell}
\author{Jonathan C. Karsch}
\author{Isaiah Gray}
\author{Gregory D. Fuchs}\email{gdf9@cornell.edu}
\affiliation{School of Applied and Engineering Physics, Cornell University, Ithaca, NY, United States.}
	
%
\begin{abstract}
Magnetic microscopy that combines nanoscale spatial resolution with picosecond scale temporal resolution uniquely enables direct observation of the spatiotemporal magnetic phenomena that are relevant to future high-speed, high-density magnetic storage and logic technologies. Magnetic microscopes that combine these metrics has been limited to facility-level instruments. To address this gap in lab-accessible spatiotemporal imaging, we develop a time-resolved near-field magnetic microscope based on magneto-thermal interactions. We demonstrate both magnetization and current density imaging modalities, each with spatial resolution that far surpasses the optical diffraction limit.   In addition, we study the near-field and time-resolved characteristics of our signal and find that our instrument possesses a spatial resolution on the scale of 100 nm and a temporal resolution below 100 ps. Our results demonstrate an accessible and comparatively low-cost approach to nanoscale spatiotemporal magnetic microscopy in a table-top form to aid the science and technology of dynamic magnetic devices with complex spin textures.
\end{abstract}
\maketitle
%
Advanced magnetic microscopies are a key tool for advancing our understanding of novel magnetic phenomena such as skyrmions, spinwaves, and domain walls \cite{SkyrmionSpatial_NC13,SpinWaves,DomainWalls}. Imaging these phenomena increasingly requires 10-100 nanometer spatial resolution \cite{Spatial_NC11,Spatial_NC12,SkyrmionSpatial_NC13,Spatial_NC14,Spatial_NC15} and 10-100 picosecond temporal resolution \cite{Temporal_NC10,Temporal_JB}. Simultaneously achieving both resolutions enables study of both static and dynamic nanoscale magnetic textures that are interesting for high-performance technology, such as skyrmion dynamics and spin torque oscillators \cite{SkyrmionDynamics1,STO_2006,STO_2017}. Unfortunately, most magnetic microscopies offer either nanoscale spatial resolution or picosecond temporal resolution, but not both. The techniques with both of these characteristics include x-ray based microscopes \cite{XRay} which require a synchrotron facility, and time-resolved electron microscopes \cite{LTEM1,LTEM2} which are expensive and not widely available. No low-cost and widely accessible tabletop techniques currently exist.

To break free of the optical diffraction limit while retaining the stroboscopic imaging capabilities offered by pulsed lasers, our approach is to use picosecond thermal pulses generated by a near-field probe-sample interaction. Recently we have demonstrated time-resolved magneto-thermal microscopy, with picosecond temporal resolution and spatial resolution determined by focused light \cite{Jason_NC,Jason_YIG,Feng_2015}. In this technique, a focused light pulse generates a microscale thermal gradient. Through the anomalous Nernst effect, the local magnetization is transduced into a voltage. This approach has proven to be useful for imaging both local static and dynamic magnetization \cite{Jason_NC}, as well as an applied current density \cite{Feng_2015}. It has also been applied to a wide range of materials beyond magnetic metals such as magnetic insulators \cite{Jason_YIG} and antiferromagnets \cite{Jason_NC,Isaiah_NiO,Isaiah_PRM,Isaiah_AM,German2} using the longitudinal spin Seebeck effect. The spatial resolution of time-resolved magneto-thermal microscopy can be further extended to the nanoscale using a scanning probe for near-field excitation. We note a parallel work to ours uses near-field magneto-thermal effects in a similar way with a continuous wave thermal gradient for detecting static magnetizations \cite{German1,German2}.

Here we demonstrate the spatiotemporal near-field magneto-thermal microscope for magnetization and applied current density imaging. We confirm the near-field character of the signal by its probe-sample distance dependence. By imaging current density around a nano-constriction that provides a well-controlled nanoscale feature, we demonstrate 100-nm-scale spatial resolution. We also verify the stroboscopic capabilities of the microscope by directly measuring a gigahertz frequency current density as a function of oscillation phase. These results provide a table-top solution to nanoscale spatiotemporal magnetic microscopy aimed at emerging complex nanoscale magnetic phenomena.
\section{Results}
\subsection{Principles of scanning magneto-thermal microscopy}
Here, we introduce the operating principles of our scanning near-field magneto-thermal microscope. We organize them into three key elements: the magneto-thermal effect and its extensions at the sample under local heating, the scanning probe that measures topography, and the near-field interaction between the illuminated tip and sample. We first discuss the principles of magneto-thermal microscopy (Fig. 1a) based on measuring thermally induced electric fields described by,
\begin{align}\label{eq:Signals}
&\bm E_{\text{ANE}}(\bm x,t) + \bm E_J(\bm x, t)\nonumber
\\
&= -N\bm m(\bm x, t)\times\bm\nabla T(\bm x, t) + \bm J(\bm x, t)\Delta\rho(\Delta T, \bm x, t).
\end{align}
To experimentally access these effects, we apply a pulsed laser, either directly focused or nano-confined by the near-field tip, to create a transient local temperature increase, $\Delta T$, and a corresponding local gradient $\bm\nabla T$. The local magnetization $\bm m$ subjected to $\bm\nabla T$ generates an electric field $\bm E_{\text{ANE}}$ through the anomalous Nernst effect (ANE) \cite{ANE1,ANE2,MTSkyrmion}, with coefficient $N$ [the first term of equation \eqref{eq:Signals}]. Throughout this paper, we study in-plane magnetization, therefore the vertical thermal gradient from the laser dominates these signals. The device ANE voltage produced from the local $\bm E_{\text{ANE}}$ is proportional to the x projection of magnetization of the sample within the thermally excited region (Fig. 1a). The second term of equation \eqref{eq:Signals} provides an extension of this technique to image current density. $\bm E_J$ arises from a current density $\bm J$ passing through the thermally excited sample volume with locally increased resistivity $\Delta\rho$. Ohm's law requires an extra electric field $\bm E_J=\bm J \Delta\rho$, which is time-dependent due to the picosecond transience of laser heating of thin metal films \cite{Feng_2015}. These two signals combine to form a voltage pulse train that we amplify and demodulate by mixing it with an electrical reference pulse train \cite{Jason_NC,Feng_2015}. The voltage is then measured by lock-in amplifiers. See Methods for more details on the electrical and optical circuits.
\begin{figure}
	\adjustbox{trim={0\width} {0\height} {0\width} {0\height},clip}%
	{\includegraphics[width=0.7\columnwidth]{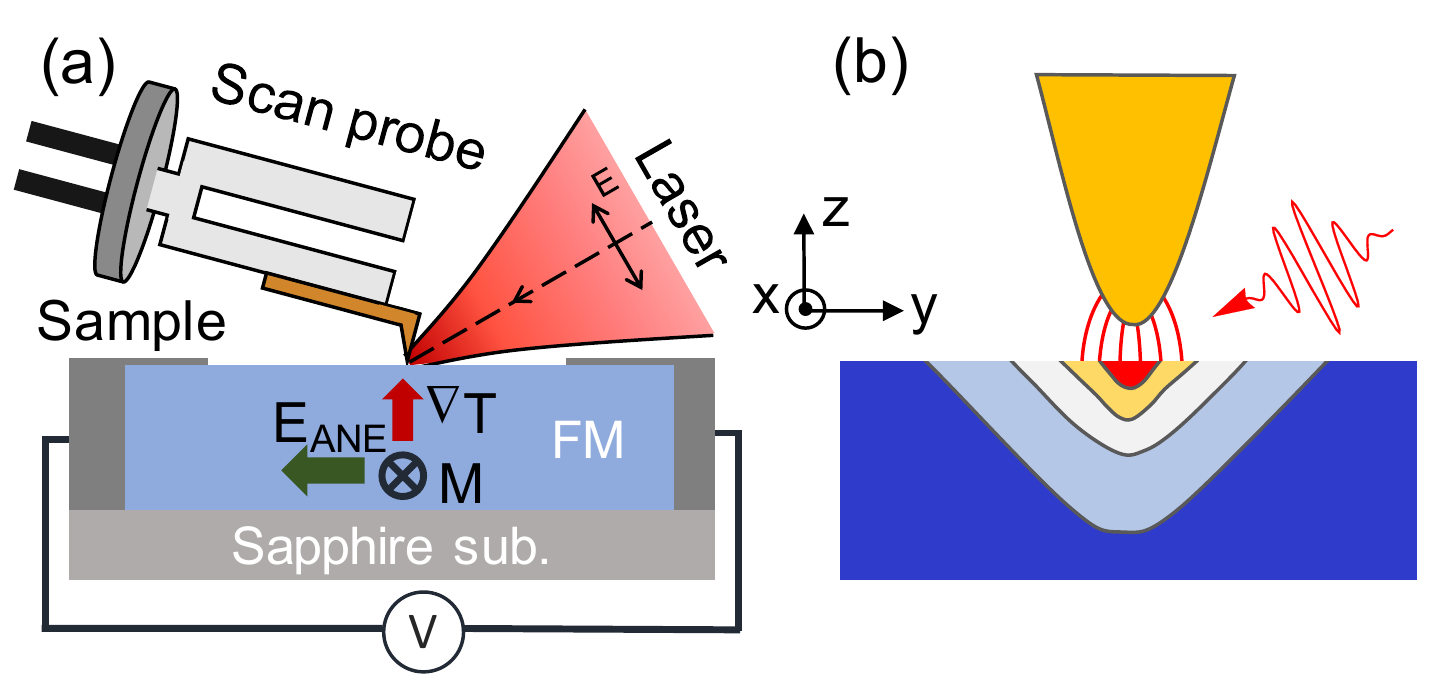}}
	\caption{\label{Figure1} Experimental schematics. (a) Schematic of the scanning near-field magneto-thermal microscope illustrating the configuration of the scan probe, the laser and the anomalous Nernst effect (ANE) in the ferromagnetic (FM) sample. (b) The near-field enhancement of the electric field generated by an optical pulse locally heats the sample in a region with size comparable to the radius of tip apex. The colors on the sample represent contours of temperature increase due to near-field heating as discussed in Ref. \cite{Jonathan}.}	
\end{figure}%

For the scanning probe, we use a tuning fork-based atomic force microscope (AFM) in tapping mode. We use probe frequency feedback to maintain an average probe-sample separation. The probe is a gold-coated Si cantilever glued to a tuning fork. The tip radius is 30 nm as received, which gets broader with scanning and reaches 50 nm or more by the time we complete the alignment procedure and record data. The probe oscillates at the tuning fork resonance frequency $f$ = 32 kHz with a typical amplitude of 40 nm. We illuminate the tip apex with laser pulses (p-polarized, angle of incidence $30^{\circ}$, laser fluence 1 mJ/$\text{cm}^{2}$ with a 76 MHz repetition rate) from a Ti:sapphire laser (3-ps-wide, wavelength $\lambda$ = 785 nm), focused using a microscope objective (numerical aperture NA = 0.42). The near-field interaction enhances the electric field in a region confined at the tip apex \cite{NF_Meng,NF_Basic,NF_Jon34,NF_Jason154}(Fig. 1b). This electric field heats the sample similarly to the focused laser, now as a nanoscale heat source. The resultant heating profile is confined to an area comparable to the tip radius, below 100 nm \cite{Jonathan}. In the measurement, the light falls onto both the sample and the tip. The focused light and the scanning probe near-field excitation both generate thermal profiles that induce ``far-field" and ``near-field" signals, respectively. We separate these contributions using lock-in detection. To recover the far-field signal, we demodulate at the 2 kHz frequency of the laser chopper. To recover the near-field signal, we use the fact that near-field interactions drop off exponentially with increasing tip-sample distance on the order of the tip radius \cite{NF_Basic}. As the probe oscillates, this nonlinearity creates a modulation at the probe frequency and harmonics (2$f$, 3$f$, \textit{etc.}). We throughout the paper demodulate the near-field signal at the second harmonic of the probe frequency. Because some far-field light can be reflected or shadowed by tip probe motion, the first harmonic contains a mixture of near- and far-field signals. The far-field contributions are linear, however, and thus they are largely suppressed in the 2$f$ demodulation channel \cite{NF_Basic,NF_Northwestern,NF_Scattering}.
\subsection{Magnetic and current measurements with scanning near-field probe}
\begin{figure}
	\adjustbox{trim={0.02\width} {0.0\height} {0.0\width} {0.0\height},clip}
	{\includegraphics[width=0.8\columnwidth]{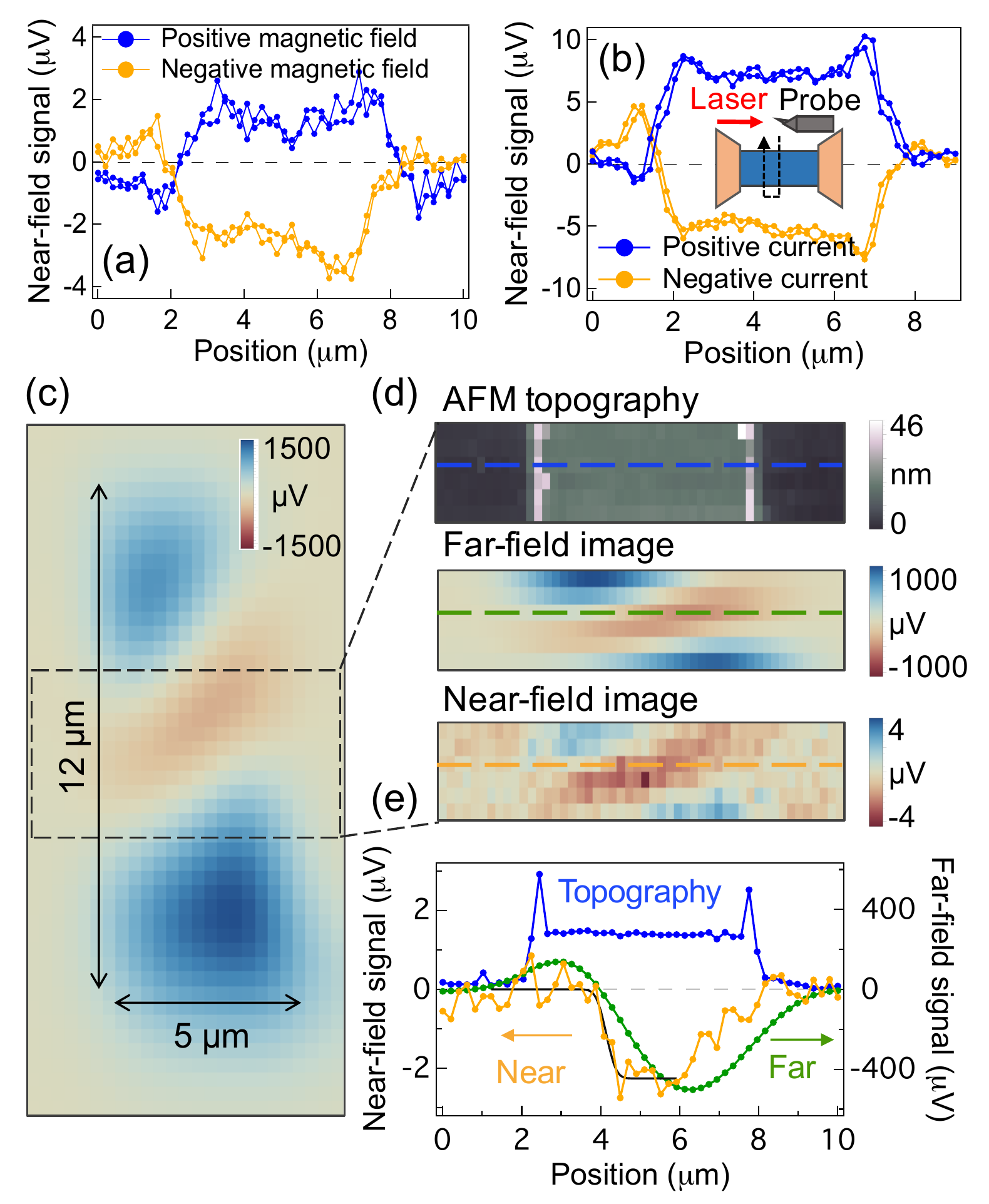}}
	\caption{\label{Figure2} Line scans and magnetic multi-domain imaging. Near-field line scans across the sample width at opposite (a) magnetization orientations and (b) applied current polarity. (a) and (b) show forward and backward line traces indicated by the inset in (b). (c) Magnetic far-field images of a multi-domain state. With a scanning probe tip, (d) images and (e) representative linecuts of topography, far-field and near-field images acquired simultaneously with the scanning probe.}
\end{figure}%
In this section, we demonstrate that a scanning near-field magneto-thermal probe can detect local magnetization and current density. We study a 5 $\mu\rm m$ x 12 $\mu\rm m$ CoFeB (4 nm)/Hf (2 nm)/Pt (4 nm) sample fabricated using photolithography \cite{Feng_2015}. A Pt capping layer as a thermal transduction layer improves uniform resolution when using different sample materials underneath \cite{Jonathan}. To align the tip and sample, we first scan the sample (on a scanning $xy$ stage) using topography in standard AFM mode. Then, with the tip retracted, we align the laser to the same features on the sample using the far-field magneto-thermal signal as in a conventional magneto-thermal microscope except here with a $30^{\circ}$ incident angle. This process coarsely aligns the laser with the tip. We then approach the tip to the surface and adjust the laser position to maximize the near-field signal. 

In Fig. 2a we show near-field imaging of a uniform magnetic state. We apply a saturating magnetic field perpendicular to the channel and measure near-field line scans across the sample width as depicted in the schematic in Fig. 2b. The near-field signal changes sign with magnetization direction, which demonstrates that we sense magnetic orientation. We note that the signal crosses zero and then decays at the sample edges rather than approaching zero sharply. This is because there are some artifact signals getting into the 2$f$ demodulation channel. The amount of artifact varies from tip to tip, and could depend on tip sharpness and the slight misalignment between the laser and tip [see Supplementary Note 2 for more discussions on artifacts]. In Fig. 2b, we apply DC current through the sample to measure near-field linecuts of current density. With the application of a 1.5 mA current, in addition to some remnant magnetic signal [Equation \eqref{eq:Signals}], the contribution from current density dominates the total signal. The near-field signal changes sign with current polarity, which confirms that the signal is sensitive to current density. 

Next, we demonstrate magnetic imaging of a multi-domain state. We demagnetize the sample using a series of minor loops with reducing field extent. First imaging without the tip, Fig. 2c shows a focused-light far-field image demodulated with respect to the optical chopper that shows magnetic domains in low resolution. In the dashed square region shown in Fig. 2c, Fig. 2d,e show images and representative linecuts of tuning fork-AFM topography, the chopper referenced far-field signal, and the probe referenced near-field signal, acquired simultaneously with the scanning probe. We see that the near-field image agrees well with the known far-field domain image, but has higher resolution. To estimate the characteristic length of the near-field features, we fit a linecut across the domain wall to $\frac{C}{2}(1+erf(\frac{x-x_0}{\sqrt{2}\delta}))$, with the full-width at half maximum (FWHM) of the curve given by $2\sqrt{2ln2}\delta$. The fit shown in Fig. 2e yields a width of 455 nm even at a $37^{\circ}$ angle with the domain wall direction, which is below the optical diffraction limit of the set-up, approximately $\lambda$/(2 NA) = 1402 nm (NA = 0.28 used for Fig. 2c-e only). However, we expect the domain wall width of this material to be wide due to its low anisotropy. The domain wall width $\delta=\pi\sqrt{\frac{A}{K_u}}$ depends on the exchange stiffness $A$ as compared to anisotropy $K_u$, the in-plane uniaxial anisotropy in this configuration. The in-plane uniaxial anisotropy of CoFeB is known to be weak or negligible \cite{Anisotropy11,Anisotropy13,Anisotropy14}. Therefore, the feature size in this sample is likely limited by the actual domain wall width rather than the instrument resolution. We further characterize instrument resolution in subsequent current imaging measurements.

Here we discuss the sensitivity of the near-field scanning probe for magnetic and current signals. For magnetic signals, we use the standard deviation of the noise in the ANE voltage $\delta_{\text{ANE}}$, and our largest size $V_{\text{ANE,sat}}$ at saturated magnetization in the 5 $\mu\rm m$ CoFeB. The magnetization angle sensitivity is calculated using $\theta_{\text{min}}=\frac{\delta_{\text{ANE}}}{V_{\text{ANE,sat}}}\sqrt{TC}$ \cite{Jason_NC}, where $TC$ is the lock-in time constant, and is estimated to be $\theta_{\text{min}}$ = $4.9^{\circ}/\sqrt{\rm Hz}$ for a 5 $\mu\rm m$-width sample. This is less sensitive than we have achieved with the conventional focused light magneto-thermal setup, however, we note that the sensitivity is dependent on several factors including the sample impedance, the Nernst coefficient, and the sample width. The signal scales inversely with the sample width due to the effective resistance shunting, and therefore the near-field scanning probe is most suitable for studying samples $\leq$ 1 $\mu\rm m$ wide. For current signal sensitivity, using the near-field signal at an applied current of 1.5 mA (Fig. 2b) and the standard deviation of the noise, we estimate the current density sensitivity to be 3.57 $\times$ $10^9$ A/($\rm m^2\sqrt{\rm Hz}$) for a 5 $\mu\rm m$-width sample. The current density signal again scales inversely with the sample width, and with better sensitivity in narrower-width devices. We note that we obtain better sensitivity to current density using the narrow channel sample discussed in subsequent current imaging measurements.
\subsection{Near-field origin of the signal}
\begin{figure}
	\includegraphics[width=0.8\columnwidth]{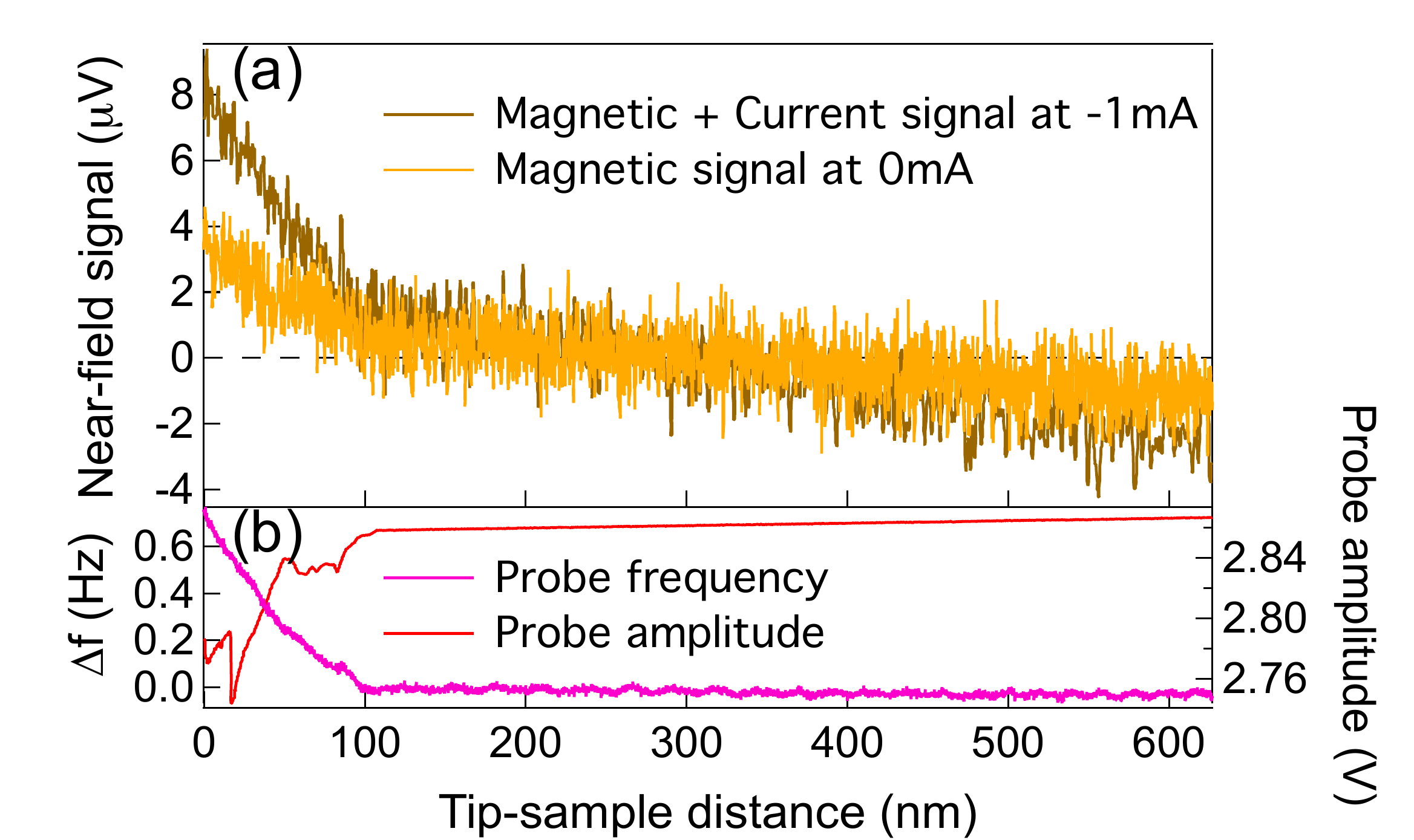}
	\caption{\label{Figure3} Near-field characteristics. Tip-sample distance dependence of (a) near-field signals and (b) probe parameters (probe frequency and amplitude).}	
\end{figure}%
Next we examine how the signal depends on tip-sample separation to study the origin of our signals. Near-field interactions are only non-negligible when the tip is at short distances from the sample, on the order of tip radius \cite{NF_Basic}. We measure both magnetic and current signals collected from the probe-demodulated lock-in as we bring the dynamically tapping tip close to the sample.  In our configuration, the laser is pre-aligned to the sample when the tip is approached, and the tip-sample displacement is controlled by retracting or approaching the tip. A near-field contribution drops off over nanoscale distances, while a far-field contribution drops off with microscale displacements corresponding to the laser intensity distribution \cite{NF_Northwestern}. We simultaneously measure the near-field signals as well as probe parameters for tip-height characterization. Figure 3 shows that the near-field signals increase when the tip is in first contact with the sample, indicated by an initial increase of the frequency and decrease of the amplitude. The 100 nm short-range increase of the signal is consistent with the near-field interaction \cite{NF_Basic}. In addition, Fig. 3a shows that the far-field artifact is largely suppressed with demodulation of the signal at 2$f$.
\subsection{Spatial resolution by current imaging}
\begin{figure}
	\includegraphics[width=0.9\columnwidth]{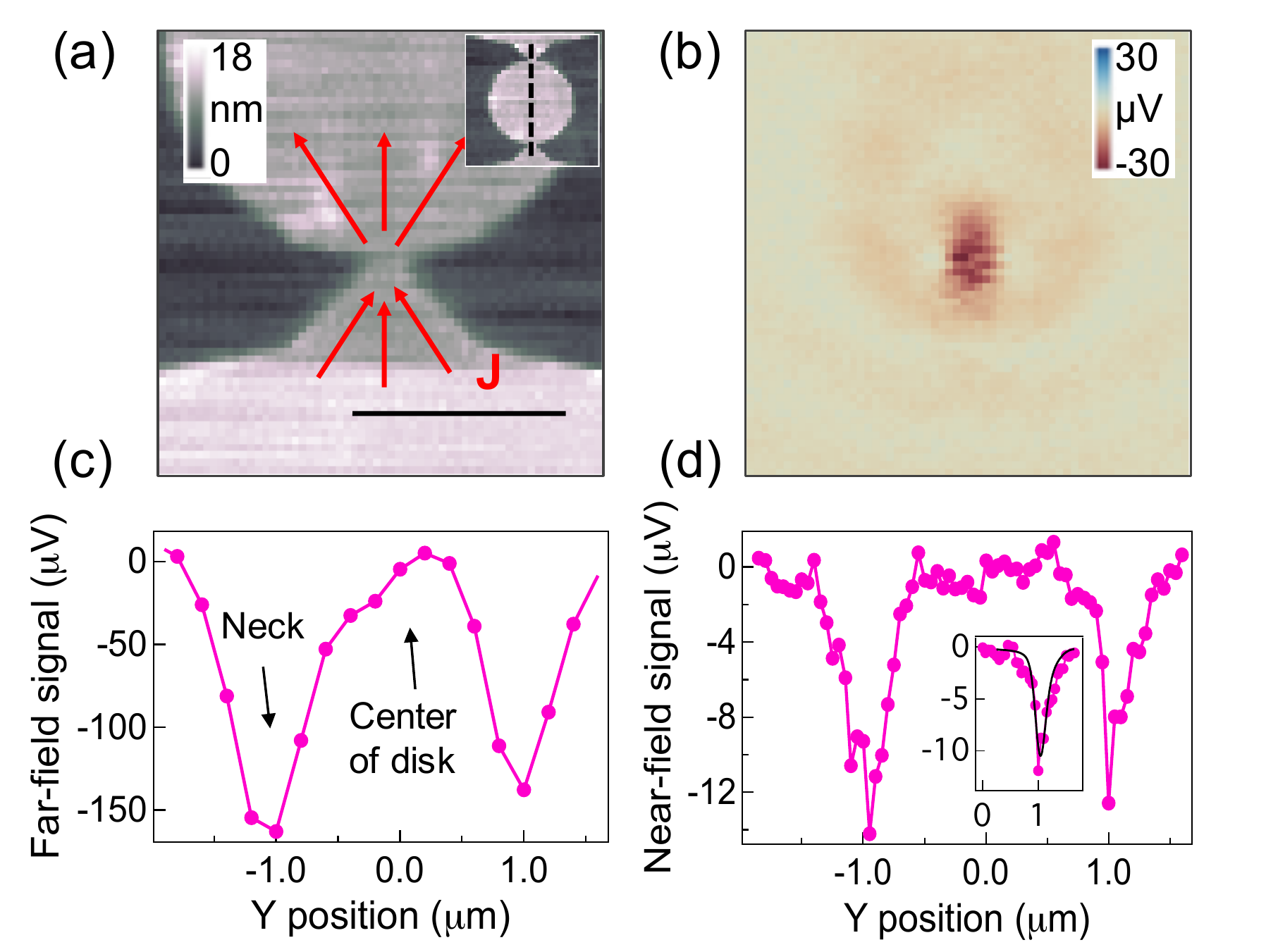}
	\caption{\label{Figure4} Current imaging and spatial resolution. (a) Topography and (b) current density images acquired by a near-field tip. The scale bar in (a) is 1 $\mu\rm m$. The inset of (a) is a wider field of view. Line cuts through two necks (as illustrated by the dashed line in the inset of (a)) of (c) far-field and (d) near-field signals for resolution comparison. (d) Inset: pure current density contribution obtained by computing the half difference of data acquired for positive and negative currents. The line is a fit to the data with width $\delta$ = 74 nm. Details of the fit are given in the Supplementary Note 1.}	
\end{figure}%
We further characterize the spatial resolution of the instrument by designing samples with sharp features suitable for benchmarking the spatial resolution. Here, we measure in current imaging mode, because current distributions can be designed and implemented through sample patterning. We use a new sample designed with narrow constrictions that confine the current density in a width comparable to the tip dimension. The sample is a $\text{Ni}_{80}\text{Fe}_{20}$ (5 nm)/Ru (2 nm) film, fabricated using e-beam lithography into a 2 $\mu\rm m$-diameter disk with two 150 nm wide necks (Fig. 4a inset). Figure 4a,b show topography and near-field current density images taken with the near-field scanning probe, at an applied current of -0.03 mA [see Supplementary Fig. 1c for data at opposite current]. The scan area on the sample is 2 $\mu\rm m$ x 2 $\mu\rm m$ and we use a lock-in time constant of 200 ms. We see that the current density is indeed concentrated at the neck. By measuring line scans through two necks (Fig. 4a inset), we compare signals between focused light far-field (Fig. 4c) and scanning probe near-field (Fig. 4d) microscopy. In Fig. 4c, we measure this sample in our conventional magneto-thermal microscope \cite{Jason_NC,Jason_YIG,Feng_2015} that uses directly focused light (angle of incidence $90^{\circ}$, numerical aperture of 0.9). By fitting the peak with a Gaussian function, the extracted FWHM resolution of 740 nm is consistent with the focused light setup resolution \cite{Jason_NC}. In Fig. 4d, the scanning near-field data has higher resolution than the far-field data. In addition, since the sample geometry is asymmetric around the neck, we expect the peak to be asymmetric, which is seen in the near-field signal.

Now we estimate a spatial resolution from Fig. 4d. Based on a sharp feature in the line scan (the left side drop-off of the right side peak), the signal makes a full scale change in voltage over $\sim$100 nm, which qualitatively gives us a resolution on the level of 100 nm. To quantitatively extract spatial resolution, we simulate the current density distribution around the neck using COMSOL [see Supplementary Note 1 and Supplementary Fig. 1a], and convolve it with a Gaussian point spread function, converted to voltage. We fit the simulated result to our data, with a Gaussian width $\delta$ as the fitting parameter. The representative fit gives $\delta$ = 74 nm (corresponding to a FWHM of 165 nm). This resolution is less than 1/4 of our focused light magneto-thermal microscopy resolution with highest NA \cite{Jason_NC}, which is consistent with the sub-diffraction resolution of near-field microscopy. In our prior work in Ref. \cite{Jonathan}, the spatial resolution for a near-field tip was simulated; for a tip radius of 45 nm (90 nm in diameter), the FWHM of the thermal gradient is 115 nm for magnetic imaging \cite{Jonathan}. Here, we note that the experimentally extracted value is for current imaging (determined by temperature increase $\Delta T$), not a magnetic measurement (determined by thermal gradient $\bm\nabla T$) \cite{Jason_NC,Feng_2015,Jonathan}. A magnetic spatial resolution is likely higher than current \cite{Jason_NC}, therefore below the extracted value. Based on the prior simulation in Ref. \cite{Jonathan}, we estimate the tip radius when we take the data to be 65 nm, which is consistent with the scanning electron microscope image of the tip taken after the measurements. We note that the resolution here is only an upper bound and that it depends on the tip sharpness at the time of scanning.
\subsection{Temporal resolution and dynamics}
\begin{figure*}
	\includegraphics[width=1.0\columnwidth]{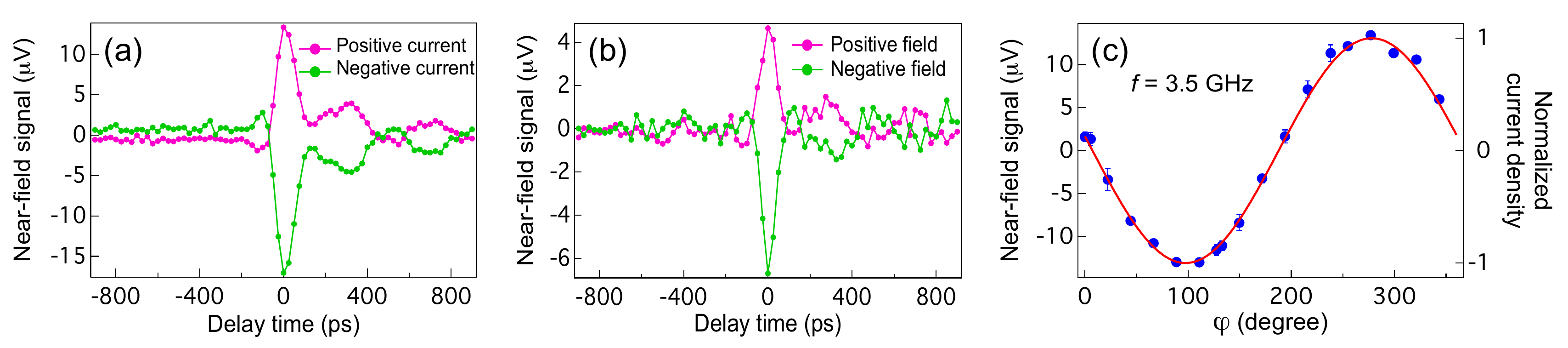}
	\caption{\label{Figure3} Stroboscopic measurements of microwave current. Time-domain measurements of the near-field voltage pulses produced by (a) current density (measured at the neck of the $\text{Ni}_{80}\text{Fe}_{20}$ sample) and (b) magnetization (measured at the center of the CoFeB sample) as a function of the pulse delay $\tau$ between the voltage pulses generated at the sample and the 100 ps reference pulses. A pulse delay of zero corresponds to the two pulse trains entering the mixer at the same time. The feature that follows the main peak is due to electrical reflections in the detection circuits \cite{Jason_NC,Jason_YIG} (c) The normalized microwave current density at the neck of $\text{Ni}_{80}\text{Fe}_{20}$ sample measured as a function of $\varphi$ stroboscopically probed by the thermal pulses. The red curve is a sinusoidal fit. The near-field signal is vertically offset to subtract the static magnetic signal contribution, obtained from the sinusoidal fitting.}
\end{figure*}%
With 3 ps laser pulses, the instrument temporal resolution to probe magnetization (current density) is determined by the temporal width of the generated thermal gradient $\bm\nabla T$ (temperature increase $\Delta T$) \cite{Jason_NC}. In prior work using focused light magneto-thermal microscopy, finite-element simulation showed that the thermal gradient pulse has a width of $\sim$10 ps. An experimental observation of magnetization resonance up to 16.4 GHz verified an upper bound of temporal resolution at or below 30 ps \cite{Jason_NC}. For thermal gradients generated by a near-field tip, the finite-element simulation gave a temporal width within 6 ps \cite{Jonathan}. Therefore, whether the excitation is from a far-field or near-field source, the picosecond-scale temporal resolution are expected to be similar. Here we experimentally examine the stroboscopic temporal properties of near-field magneto-thermal microscopy.

Figure 5a,b show a measurement of the picosecond-scale voltage pulses that are directly created in response to a transient $\Delta T$ or $\bm\nabla T$ in the presence of an applied current density or a static magnetization \cite{Jason_NC}. The short-lived thermal excitation of the sample is the origin of the stroboscopic time dependence of magneto-thermal microscopy. We measure the pulses by combining them in an electrical mixer with 100 ps electrical reference pulses that are synchronized with the laser but have a controllable delay $\tau$. The mixer output represents a time-domain heterodyning of the $V_{\text{ANE}}$ pulses, where we detect the component mixed down to DC but still containing the slower lock-in modulation \cite{Jason_NC,Feng_2015}. Figure 5a,b show the mixed output voltage as a function of the delay $\tau$, which can be understood as the temporal convolution signal of the voltage pulses with the reference pulses \cite{Jason_NC}. In both plots, the convolved signal widths are roughly 100 ps, similar to the reference pulse width. Therefore, the voltage pulses generated by the near-field thermal excitations must be shorter than 100 ps, demonstrating a time-resolved nano-probe. 

In Fig. 5c, we also demonstrate a stroboscopic capability of the scanning probe to measure current density with temporal resolution that significantly exceeds the oscillation period of a 3.5 GHz current applied directly to our device. To make this measurement, we synchronize the laser and microwave current by choosing a frequency at an integer multiple of the laser repetition rate such that the thermal pulses constantly probe the current at the same phase, $\varphi$ \cite{Jason_NC,Feng_2015}. Here, $\varphi$ is a relative phase between the microwave current and the laser with a phase offset that depends on frequency and initial conditions. Time-resolved measurements of current imply that we can use our near-field microscope to make phase-sensitive observations of microwave current within a nanoscopic volume. Fig. 5c shows normalized microwave current density as a function of $\varphi$, showing the phase-sensitive response. As we rotate $\varphi$, we observe the expected sinusoidal response in the near-field signal.
\section{Discussion}
In conclusion, we demonstrate scanning near-field magneto-thermal microscopy of magnetization and current density using picosecond thermal pulses. This work represents an important milestone for low-cost, table-top magnetic microscopy that unifies nanoscale spatial resolution with picosecond temporal resolution in the same instrument. Here we experimentally image magnetic domains with nanoscale resolution, and we verify the near-field origin of both magnetic and current density contrast with a spatial resolution on the scale of 100 nm. Additionally, we demonstrate picosecond-scale temporal resolution of current density and magnetization, enabling a stroboscopic nanoscale probe of emerging magnetic devices in which we can directly probe both a stimulus and its response. For example, nanoscale phase-sensitive imaging could be useful for understanding mechanisms and phase-relationships in mutual synchronization locking of spin-orbit torque oscillators \cite{STO_2017}. Meanwhile, the capability to image current from DC to microwave frequencies could clarify the origins of magnetic resonances in spatially non-uniform spin-Hall nano-devices \cite{Feng_2016}.

We are optimistic that scanning probe magneto-thermal microscopy can be further developed into a powerful tool to study the dynamics of nanoscale magnetic devices and spin textures. For example, the spatial and temporal resolution of our microscopy is compatible with measuring magnetic resonance from a single magnetic skyrmion, rather than from an ensemble \cite{SkyrmionDynamics1}.  Furthermore, magneto-thermal microscopy is compatible with a wide palette of magnetic materials beyond magnetic metals, including ferrimagnetic magnetic insulators \cite{Jason_YIG} and antiferromagnetic insulators \cite{Isaiah_NiO} (via the spin Seebeck effect), thus offering a versatile imaging tool.  Scanning probe magneto-thermal microscopy is especially suitable for devices with sub-micron channel widths, and using materials that have strong magneto-thermal effects. Indeed, improving the sensitivity of this instrument will be the most important step forward in its future development.  The magnetization sensitivity can be further improved through optimization of the instrument (\textit{e.g.} detection electronics, impedance matching, probe amplitude optimization \cite{NF_Basic}, tip sharpness) and through incorporation of near-field engineering (including a plasmonic coupling grating \cite{NF_Northwestern,NF_Jon25,NF_Jason152,NF_Jason155}).
\section{Methods}
\subsection{Optical excitation of near-field scanning tip}
\begin{figure}
	\includegraphics[width=0.7\columnwidth]{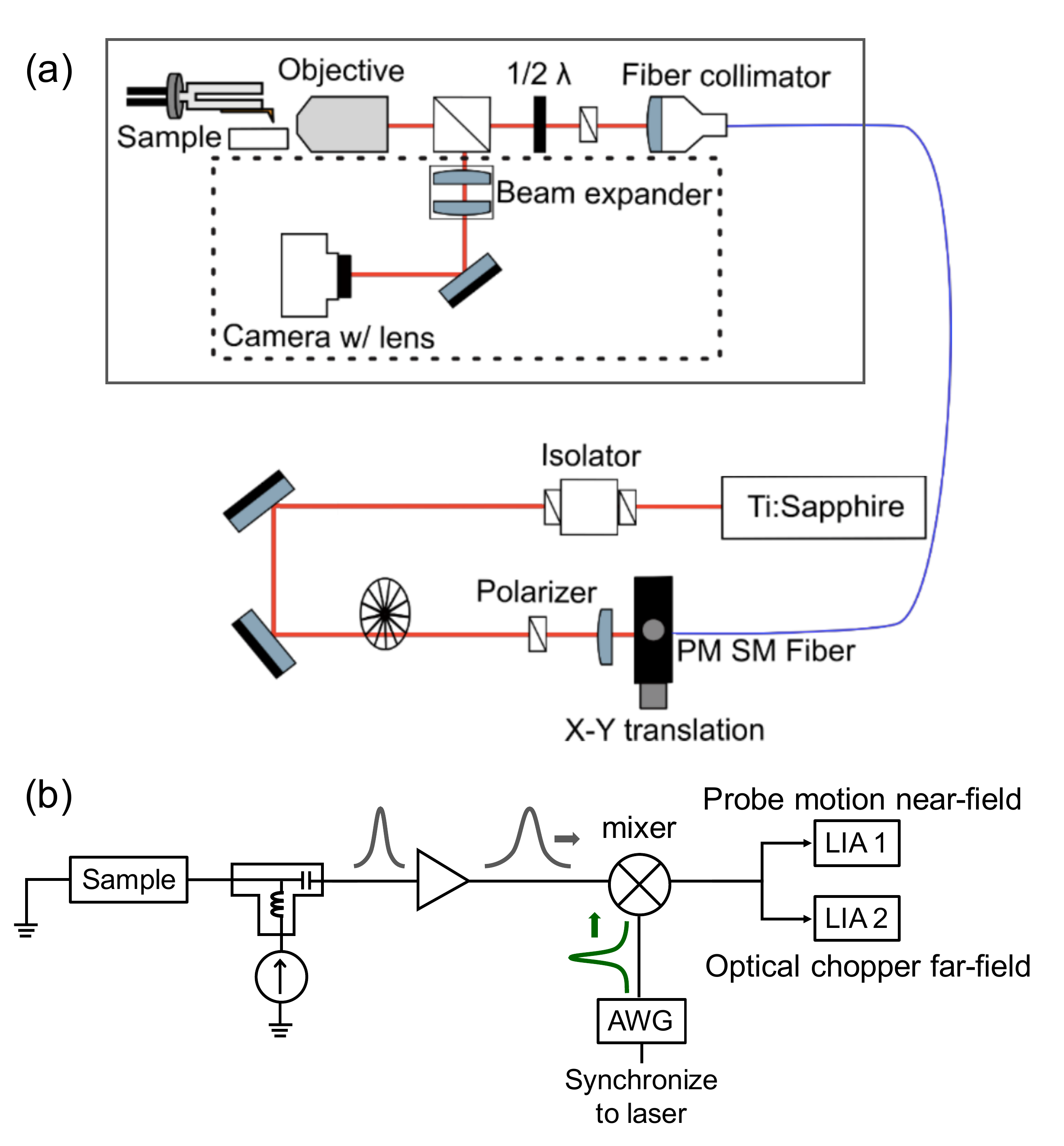}
	\caption{\label{Figure6} Optical setup and electrical circuits. (a) Schematic of the optical path used in the scanning magneto-thermal microscopy. The outer solid boxed area outlines the enclosure. The inner dashed boxed area is the optical microscope used to image the tip and sample through the objective. The objective, beam splitter, half-waveplate, polarizer, and fiber collimator are mounted on a $xyz$-translation stage at an angle of $30^{\circ}$ to the sample surface. (b) Schematic of electric circuits for detection in scanning magneto-thermal microscopy.}	
\end{figure}%
Figure 6a shows the optical path used to illuminate the probe tip in our scanning magneto-thermal microscope. A Ti:sapphire laser (Coherent MIRA 900) generates 3 ps pulses with a 76 MHz repetition rate and a 785 nm wavelength. The light first passes through a Faraday isolator, a mechanical chopper (2 kHz), and a polarizer before being coupled into a polarization maintaining (PM) single mode (SM) fiber. The fiber carries the light into the enclosure of the scanning magneto-thermal microscope. The light that emerges from the fiber is collimated and sent through another polarizer and waveplate combination to ensure that the light is p-polarized. We then focus the laser onto the scanning probe tip apex using a microscope objective with a 20 mm working distance and 0.42 NA. A beam expander and camera are used to image the probe set-up to enable alignment of the laser, tip, and sample.
\subsection{Electrical detection}
The electrical circuits are the same as those of the focused light magneto-thermal microscope \cite{Jason_NC,Feng_2015}. The voltage pulses generated at the sample are collected into a microwave transmission line and amplified by 40 dB with 3 GHz bandwidth. The amplified pulses are then detected by mixing them in a DC-12 GHz mixer with 1.2 ns-duration electrical reference pulses, generated by an arbitrary waveform generator (AWG) that is referenced to the laser repetition rate. (For data in Fig. 5a,b, different amplifiers with 15 GHz bandwidth and pulses of 100 ps width - the shortest reference pulses that can be produced by the AWG - are used for the temporal upper bound characterization.) This technique demodulates the pulsed signal at the laser repetition frequency. This approach is similar to the homodyne detection used in a lock-in amplifier (LIA), except instead of using a single frequency sinusoidal reference signal, here we use a synchronous pulse train to demodulate the sum of multiple harmonics of laser repetition frequency, therefore recovering a larger demodulated signal. The voltage is sent to lock-in amplifiers which demodulate the signal at probe motion and chopper frequencies, respectively.
\section{Acknowledgments}
We thank Dr. Long Ju, Dr. Samuel Berweger, and Harry Cheung for helpful discussions. Time-resolved and current imaging studies were supported by the DOE Office of Science, Basic Energy Sciences (DE-SC0019997). Preliminary development and static magnetic imaging was supported by the AFOSR (FA9550-14-1-0243).  This work made use of the Cornell Center for Materials Research Shared Facilities which are supported through the NSF MRSEC program (DMR-1719875), and the Cornell NanoScale Facility, a member of the National Nanotechnology Coordinated Infrastructure (NNCI), which is supported by the National Science Foundation (Grant NNCI-2025233).
\section{Author contributions}
G.D.F. and J.M.B. conceived the idea of the microscopy. J.M.B., J.C.K. and C.Z. built the apparatus. C.Z. and J.M.B. developed the instrument and acquired the data with assistance from I.G. C.Z. designed, fabricated the samples and performed numerical simulations. C.Z. and G.D.F. prepared the manuscript and all authors revised the manuscript.  

\newpage
\renewcommand{\figurename}{Supplementary Figure}
\setcounter{figure}{0}
\begin{center}
	{\large \textbf{Supplementary Information\\Nanoscale magnetization and current imaging using scanning-probe magneto-thermal microscopy}}
\end{center}
\section{Supplementary Note 1: Current density distribution and spatial resolution extraction}
\subsection{A. Current density distribution}
We simulate current density distribution around the neck (the tapered junction at the base of the $\text{Ni}_{80}\text{Fe}_{20}$ disk) for quantitatively extracting the resolution. The neck has a 150 nm width, as shown in the atomic force microscopy (AFM) image in Supplementary Figure 1(a) inset. We set the sample geometry in COMSOL closest to the real AFM profile [Supplementary Figure 1(a) inset]. We apply a current of 0.03 mA in the simulation, and Supplementary Figure 1(a) shows the image of the simulated $y$ component of current density.
\subsection{B. Simulation equation for current density signal contribution}
The current density signal contribution comes from the laser-induced local heating. The local resistivity increase $\Delta\rho$ from the local temperature increase $\Delta T$ times the applied current density $J$ produces an additional voltage as described in the main text. To simulate this signal, we consider the linear response regime of the resistivity dependence on temperature, such that $\rho(T) = \rho_0(1+\alpha\Delta T)$, \textit{i.e.} $\Delta\rho=\rho_0\alpha\Delta T$, where $\rho_0$ is the resistivity at room temperature, $\Delta\rho$ is the resistivity increase, and $\alpha$ is the temperature coefficient of resistivity. To derive the total voltage measured at the electrical contacts, we follow Ref. \cite{Jason_NC} and use a simplified resistor model. We subdivide the sample into blocks labeled by $k$, with dimensions of $dx$, $dy$ and $dz$. We consider the laser heating-induced additional voltage as our local voltage source. Therefore, $\Delta V_k\approx i_k\Delta R=J\Delta\rho dy=J\rho_0\alpha\Delta T dy$, where $\Delta V_k$ ($i_k$) is local additional voltage (current) in block $k$, here $y$ is the sample length direction, \textit{i.e.} the direction of the linecut in Figure 4(d), and $x$ is the sample width direction. Using Kirchhoff's laws, the global additional voltage at sample length $\Delta V_J$ calculated from the local voltage to take into account resistors in parallel is given by,
\begin{equation}
	\Delta V_J = \rho_0\alpha\int\frac{J(x,y)\Delta T(x,y_{\text{tip}} - y)}{w(y)} dx dy
\end{equation}
where we simulate the local current density $J(x,y)$ using COMSOL, represent the laser-induced temperature profile $\Delta T(x,y_{\text{tip}}-y)$ as a Gaussian function at tip location $(0,y_{\text{tip}})$, and get the width $w(y)$ from AFM topography profile which we also use in the COMSOL simulation. We treat $J$ and $\Delta T$ as uniform in the $z$ direction for simplicity.
\subsection{C. Quantitative extraction of spatial resolution from current imaging}
\begin{figure}
	\includegraphics[width=0.6\columnwidth]{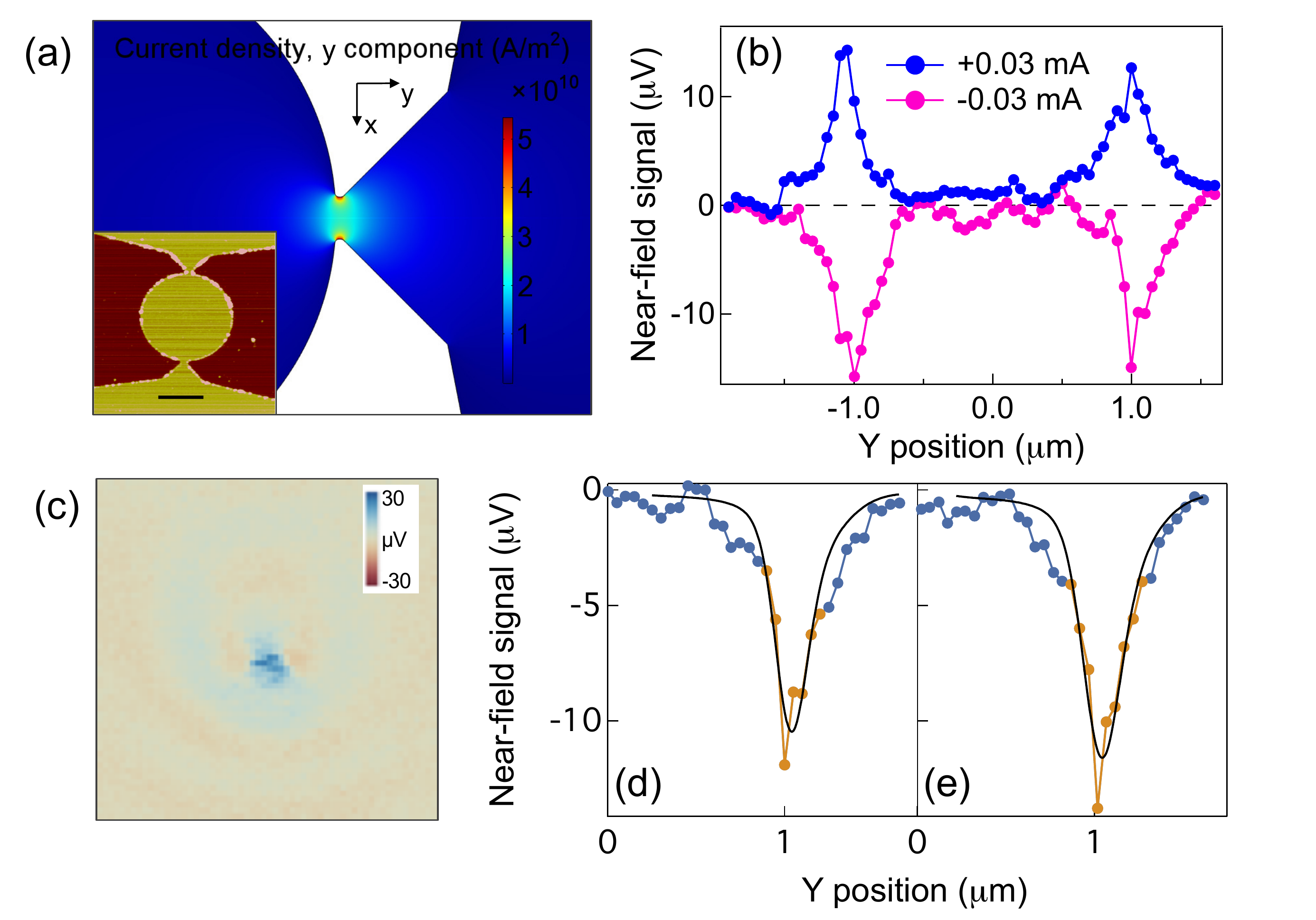}
	\caption{\label{FigureS3} Current density distribution and spatial resolution extraction. (a) COMSOL simulation on $y$ component of current density. The coordinate system corresponds to the current density image, not the inset. Inset: AFM image of the device. The scale bar is 1 $\mu\rm m$. (b) Example current density linecuts through two necks at both current polarities. (c) Current density image acquired by a near-field tip at +0.03 mA (-0.03 mA is presented in the main text). Simulated curve fitted to the right side peak of half difference data, with a Gaussian width (d) $\delta$ = 74 nm and (e) $\delta$ = 88 nm.}	
\end{figure}%
To quantitatively extract the resolution for imaging current density, we numerically convolve the current density from the COMSOL simulation with a Gaussian point spread function of width $\delta$, converted to voltage. We measure line scans across the top and bottom $\text{Ni}_{80}\text{Fe}_{20}$ necks at positive and negative currents [Supplementary Figure 1(b)], and take the half difference of the two to subtract out the magnetic contribution. We use the half difference because our simulation is for the current density signal only. We note that the individual linecuts are sharper than the half difference, and there could be some loss of resolution due to any Y position shift between the two linecuts. We fit the simulated result to the right side peak of the half difference data to extract $\delta$. We notice that when the right side peak drops off to close to zero, there is an extra contribution that might be an artifact, and that leads to a broader slope. We fit to the data range indicated in orange points in Supplementary Figure 1(d), to balance the overall fit and the fit to the top of the peak which is less sensitive to the artifact. In this case, we find a Gaussian width of $\delta$ = 74 nm. We also show the fit to another parallel linecut that is 50 nm apart in x-direction, and we find the $\delta$ = 88 nm [Supplementary Figure 1(e)].
\subsection{D. Further discussions on spatiotemporal profiles of $\bm\nabla T$ and $\Delta T$}
Here, we provide further discussion about the spatiotemporal profiles of a temperature gradient $\bm\nabla T$ and a temperature increase $\Delta T$, which have been simulated and studied in prior work \cite{Jason_NC}. First, we explain the spatial resolution comparison between probing magnetization and current density. When we measure in-plane magnetization, the spatial resolution is limited by the areal extent of the z-component of $\bm\nabla T$, integrated over the time during which $\bm\nabla T_z$ is nonzero. Even though $\Delta T$ spreads laterally, $\bm\nabla T_z$ does not spread much beyond the excitation spot before it collapses to uniform temperature through the metal film \cite{Jason_NC}. Therefore, the spatial resolution of the magnetic signal (limited by the excitation size) should be higher than that of the current signal (limited by the lateral spreading of $\Delta T$). Second, we discuss the complications of temporal response for current density measurements. Unlike the temporal profile of $\bm\nabla T$ for magnetic signals, which is a sharp pulse with $\sim$10 ps width, the temporal profile of $\Delta T$ is highly asymmetric, with a sharp initial drop-off within 100 ps, and then a slowly decaying tail extended to hundreds of picoseconds \cite{Jason_NC}. In a high-frequency microwave current measurement, the sharpest first 100 ps of $\Delta T$ dominates the temporal performance and gives the high-frequency responses \cite{Feng_2015}, with the tail creating a partial cancellation in the signals.
\section{Supplementary Note 2: Tip alignment and tip artifact}
An important step in operating the scanning probe magneto-thermal microscope is to align the tip apex to the position in 3D space that corresponds to the point at the center of the laser focus while the tip is engaged to the sample. The laser is oriented with its central axis at $30^{\circ}$ relative to the sample surface. The microscope objective is on a $xyz$ stage controlled with piezo motors to adjust the laser position with $<$ 30 nm precision for alignment. During the imaging, the sample is on a $xy$ stage for scanning, and the probe tip is on its own z stage for the height feedback. We first align the tip and sample by topography scanning to park the tip on top of a desired location. We coarsely focus the laser onto the sample at the same desired location by optimizing the far-field signal when the tip is retracted. We then fine-align the tip, laser and sample all together by maximizing the near-field signal in 2$f$ when the tip is in contact with the sample. We note we also observe near-field signals at 3$f$ and 4$f$.

We demodulate at the probe frequency and harmonics for the near-field signal. If some far-field light, either directly illuminated or scattered by the tip shaft onto the sample, gets modulated by the probe motion, it can possibly go into probe demodulated channels as an artifact. An example is the tip shadowing artifact. With the tip oscillating up and down, part of the light will be unblocked and blocked by the tip, as shown in the Supplementary Figure 2(b). Since the length scale of far-field wavelength is much larger than the probe amplitude, far-field artifact changes approximately linearly with the probe motion \cite{NF_Northwestern}. Therefore, it appears in the 1$f$ channel but it is largely suppressed at 2$f$ channel \cite{NF_Northwestern}.
\begin{figure}
	\includegraphics[width=0.6\columnwidth]{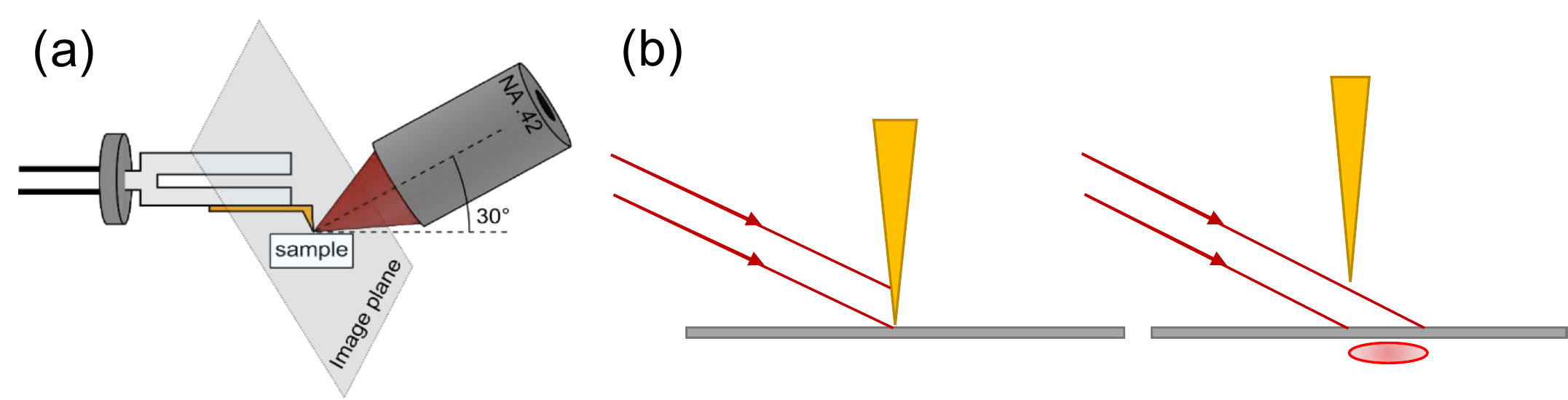}
	\caption{\label{FigureS4} Tip alignment and tip artifact. (a) Schematic representation of the alignment between the objective, scanning probe, and sample. The plane imaged by the microscope objective is tilted with respect to the sample. (b) Schematics of tip shadowing artifact.}	
\end{figure}%


\end{document}